\documentclass[11pt]{article}

\usepackage{amsmath}
\usepackage{amssymb}
\usepackage[russian,french,greek,english]{babel}
\usepackage{caption}
\usepackage{colortbl}
\usepackage{enumitem}
    \setlist{nosep,topsep=3pt,partopsep=5pt}
\usepackage{fancyvrb} 
\usepackage{float}
\usepackage[stable,bottom]{footmisc}
\usepackage{footnote}
\usepackage{geometry}
\usepackage{graphicx}
\usepackage{hyperref} 
\usepackage{import}
\usepackage[utf8]{inputenc}
\usepackage[framemethod=TikZ]{mdframed}
\usepackage{multirow}
\usepackage{tabularx}
\usepackage{textcomp}
\usepackage{tikz}
    \usetikzlibrary{patterns}
\usepackage[normalem]{ulem}
\usepackage{xcolor}
\usepackage{wrapfig}

\mdfsetup{
    skipabove=6pt,
    skipbelow=6pt,
    innertopmargin=12pt,
    innerbottommargin=12pt,
    topline=false,
    bottomline=false,
    leftline=false,
    rightline=false,
    backgroundcolor=gray!9,
    roundcorner=10pt
}

\newenvironment{extmdframed}
{
    \begin{mdframed}
    
} {
    \end{mdframed}
}

\newcommand{\abs}[1]{\left\vert#1\right\vert}

\newcommand{\mylittlelinebreak}{\hfill\vspace{2mm}}

\def \CC {{\mathbb{C}}}

\def \ZZ {{\mathbb{Z}}}

\definecolor{colorcomment}{rgb}{0.8,0,0}
\definecolor{gray1}{rgb}{0.1, 0.1, 0.1}
\definecolor{gray2}{rgb}{0.2, 0.2, 0.2}
\definecolor{gray3}{rgb}{0.3, 0.3, 0.3}
\definecolor{gray4}{rgb}{0.4, 0.4, 0.4}
\definecolor{gray5}{rgb}{0.5, 0.5, 0.5}
\definecolor{gray6}{rgb}{0.6, 0.6, 0.6}
\definecolor{gray7}{rgb}{0.7, 0.7, 0.7}
\definecolor{gray8}{rgb}{0.8, 0.8, 0.8}
\definecolor{gray9}{rgb}{0.9, 0.9, 0.9}
\definecolor{colortranslatornote}{rgb}{0.0,0,0.8}

\begin{document}

\title{On integer sequences for rendering limit sets of Kleinian groups}
\author{Alessandro Rosa}

\maketitle

\begin{abstract}
We present a technique for rendering limit sets for kleinian groups, based upon the base transformation of integers and which aims at saving memory resources and being faster than the traditional dictionary based approach.
\end{abstract}

\section{The framework}\label{section_env}
We are going to work with linear fractional transformations (LFT) in the form
\begin{equation}\label{eq_01}
z_1=g_n(z)=\frac{a_nz_0+b_n}{c_nz_0+d_n},\hspace{0.5cm}a_n,b_n,c_n,d_n,z_0\in\CC,\hspace{0.5cm}ad-bc\neq 0,\hspace{0.5cm}n\in\ZZ
\end{equation}
\noindent known as \emph{Möbius transformation} (or \emph{map}) after the in-depth studies carried out by August Ferdinand Möbius (1790--1868) during the beginnings of the XIX century. These maps build the set $\mathcal{M}$, also known as $Aut(\hat{\CC})$, of automorphisms $g_n$ of the complex plane $\CC$. Inversion, $\displaystyle g^{-1}_n(z)=\frac{d_nz-b_n}{-c_nz+a_n}$, and composition, $g_1\circ g_2=g_1(g_2)=g_3$, are closed in $\mathcal{M}$. The latter operation enjoys the same properties as of $2\times 2$ matrix, namely
$$\begin{pmatrix}a_1 & b_1\\c_1 & d_1\end{pmatrix}\circ\begin{pmatrix}a_2 & b_2\\c_2 & d_2\end{pmatrix}=\begin{pmatrix}a_1a_2+b_1c_2 & a_1b_2+b_1d_2\\a_2c_1+c_2d_1 & b_2c_1 + d_1d_2\end{pmatrix},$$
and the associative property: $(g_1\circ g_2)\circ g_3=g_1\circ(g_2\circ g_3)$. \label{par_no_primitive_elements}This last relation naturally extends to every chain
\begin{equation}\label{eq_02}
g_1\circ g_1\circ g_2\circ g^{-1}_1\circ g^{-2}_1\circ g_2\circ \dots\ .
\end{equation}
No transformation in $\mathcal{M}$ can be regarded as primitive because of being decomposable into such formulas. In addition, the identity $I(z)=z=g\circ g^{-1}$ represents a special Möbius map whose coefficients are $a=d=1$, $b=c=0$ and it plays the role of neutral element in $\mathcal{M}$: $g_n\circ I=I\circ g_n=g_n$. The inverse $g^{-1}$ commutes with $g$: $g\circ g^{-1}=g^{-1}\circ g=I$.\footnote{The composition of these transformations is not generally commutative anyway.} Playing with the four coefficients also shows that LFT could turn into any elementary transformation of the plane: rotation $e^{2\pi i\theta}z$ (for $b=c=0$, $d=1$, $a=e^{2\pi i\theta}$ for $\theta\in[0,2\pi)$), translations $az+b$ ($c=0$, $d=1$), inversions $b/z$ ($a=d=0$, $c=1$), contractions or dilations $az$ ($b=c=0$, $d=1$) for $\abs{a}<1$ or $\abs{a}>1$ respectively.

The number of fixed points $\lambda$ of $g$, satisfying the relation $\lambda=g(\lambda)$, is at most two and are computed by
$$z_{\pm}=\frac{(a-d)\pm\sqrt{(d-a)^2}+4bc}{2c}.$$
Here the expression $tr(g)= a+d$, defined \emph{trace}, shows up to be an essential tool for determining the position and the number of such points as well as, more extensively, for classifying all LFTs under these four categories: elliptic, parabolic, hyperbolic and loxodromic, whether $0\leq tr^2(g)<4$, $tr^2(g)=4$, $tr^2(g)>4$ and $tr^2\in\CC\backslash[0,4]$ respectively; parabolic transformations have two identical fixed points, whereas they are distinct in all other cases.

We have gathered enough information to state that $\mathcal{M}$ introduces a variegated scenario and that it is the largest \emph{algebraic group} of Möbius maps: this variety will produce sub\-groups $\mathcal{G}$ of $\mathcal{M}$, that en\-joy spe\-ci\-al pro\-per\-ti\-es according to the nature of the relations between the coefficients in (\ref{eq_01}), often ruled by the trace operator. The theory of these sub\-groups roots to the concept of \emph{generators}, being LFTs that are \emph{conventionally assumed} to give rise to all other elements in $\mathcal{G}$: in fact, since they are always decomposable into other LFTs, the generating role is apparent and their formulas are just intended to give concrete example of the special relations satisfied by the coefficients within.

The action of $\mathcal{G}$ is defined as \emph{freely discontinuous} when $g(U)\cap U=\emptyset$ holds, for $z\in U\subset\CC$ \cite[p. 16]{Maskit-1988}. The combination of (\ref{eq_02}) with (\ref{eq_01}) yields an ordered sequence of image values, $z_1$, $z_2$, \dots, $z_n$, which is defined as \emph{orbit} in the theory of dynamical systems.\footnote{Iteration is a special case of chains (\ref{eq_02}) including one only transformation: $g_1\circ g_1\dots g_1\circ \dots\ $.} It is proven that these orbits are \emph{asymptotically stable} \cite[p. 17]{Maskit-1988}; it is then natural to get interested into $(1^\circ)$ the final destination, i.e., the \emph{limit value}, of collectively taken, the \emph{limit set} $\Lambda$. Like most limit sets, even $\Lambda$ is asymptotic and then not algorithmically feasible, because of involving infinitely many combinations (\ref{eq_02}). The orbits shall be necessarily halted at some finite step $d<\infty$, so that we just deal with approximations $\Lambda_d$ of $\Lambda\equiv\Lambda_\infty$, assuming $\displaystyle\lim_{d\rightarrow\infty}\Lambda_d=\Lambda_\infty\equiv\Lambda$.

We just need these basic notions in what follows. For further and in-depth information, refer to these bibliographic sources \cite{Beardon-1983, Maskit-1988, Indras-2002}.\mylittlelinebreak

\subsection{Inversions and isometric circles.}
\begin{figure}[!b]
\centering
\begin{tabular}{cp{1cm}c}
\input{figs/pic/circle_inversion.pic}
& &
\includegraphics[height=3.4cm]{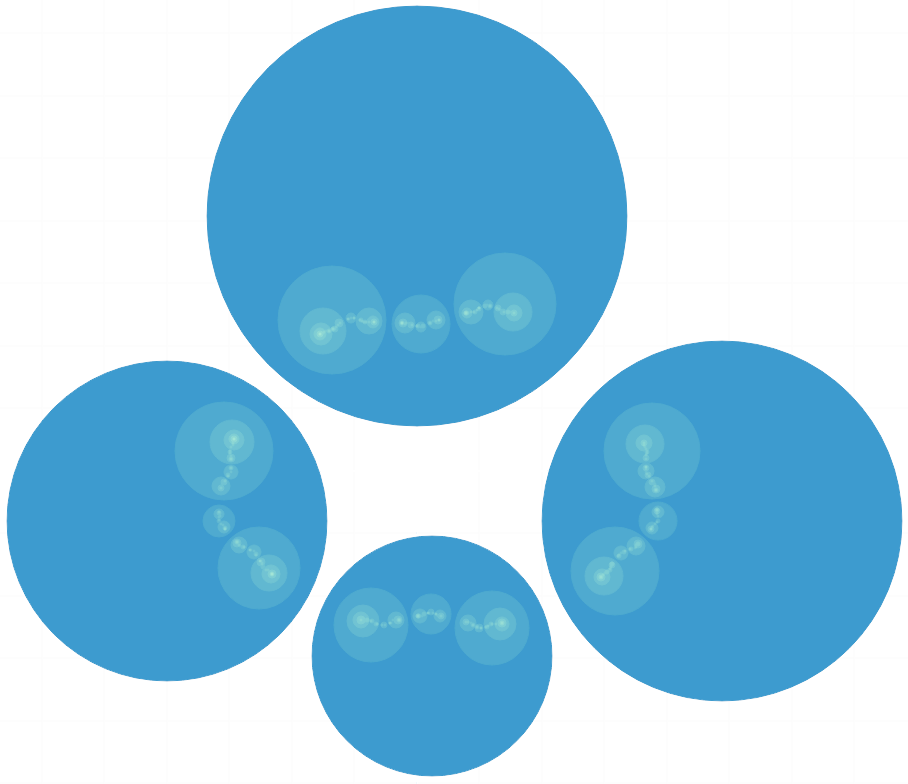}\\
\end{tabular}
\caption{\textbf{Tessellation by circle inversions.} The top diagram il\-lu\-stra\-tes how cir\-cle in\-ver\-si\-on works.  At the bottom, the disc images under the action of a subgroup whose generators have no self-intersecting inversion circles and known as of \emph{Schottky} type.}\label{fig_circle_inversion}
\end{figure}

We will see that circles are special shapes in the theory of Möbius maps. There exists a subgroup $G$ of Mobius maps
\begin{equation}\label{eq_03}
T(z)=a_C+\frac{r_C^2}{\overline{z-a_C}},
\end{equation}
which are defined \emph{circle inversions} (or \emph{reflections}) and map circles $C$ to themselves (so they are conformal transformations), whereas interior and exterior are swapped (fig. \ref{fig_circle_inversion}/a). Sizes of the objects inside of these regions would not be preserved.

That said would be enough to guess that the repeating application of circle inversions could generate sequences of image objects whose size would progressively decrease for instance (fig. \ref{fig_circle_inversion}); hence we could speak of limit sets here in the previous terms. For instance, let $C$ be a circle centered at $a_C=x_C+iy_C$ and with radius $r_C$. Conversely, we can determine $a_C$ and $r_C$ in (\ref{eq_03}) from $C$. Hence $T$ maps every object outside (resp. inside) $C$ inside (resp. outside) it (fig. \ref{fig_circle_inversion}); then, $T\circ T=I$, or $T^2(z)=I$. $C$ is defined \emph{inversion} circle and it is the inverse of itself, i.e., \emph{invariant} under $T(z)$. From now on, let this circle be denoted as $C_{\textnormal{INV}}$.

There exists another family of invariant circles for (\ref{eq_01}), which are called \emph{isometric},\footnote{This word comes from the union of the two Greek terms \emph{iso} and \emph{metric} = \emph{same size}.} where $g(C_{\textnormal{ISO}})=C_{\textnormal{ISO}}=g^{-1}(C_{\textnormal{ISO}})$  and which are algebraically defined by the equality $\abs{cz+d}=1$, with $c\neq 0$.\footnote{They are centered at $-d/c$ and with radius $1/\abs{c}$, hence there are no such circles when the LFT is not a rational map.} Isometric circles will be denoted $C_{\textnormal{ISO}}$ here.

\begin{figure}[!h]
\centering
\begin{tabular}{cp{0.1cm}cp{0.1cm}c}
    \includegraphics[height=3.6cm]{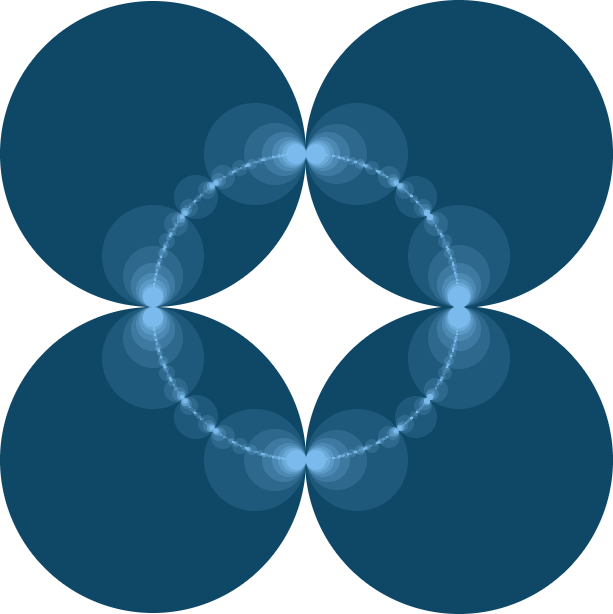}
    & &
    \includegraphics[height=3.6cm]{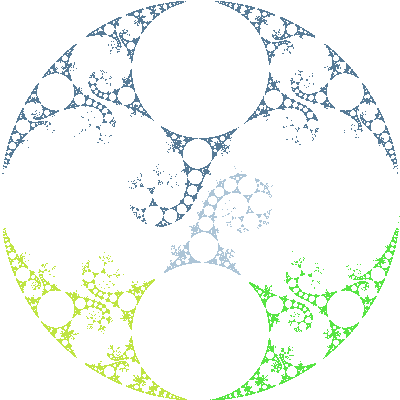}
    & &
    \includegraphics[height=3.6cm]{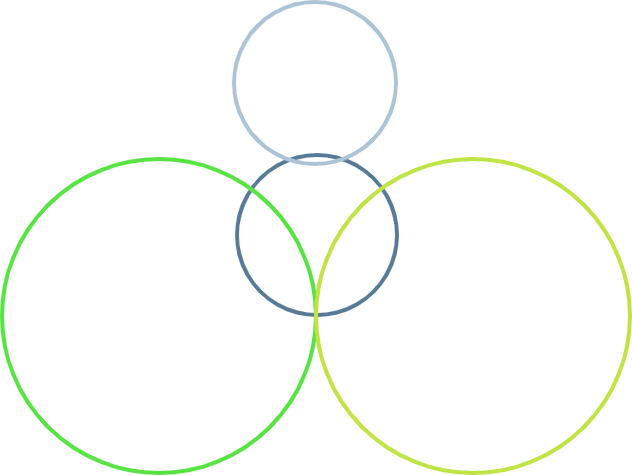}
    \\
    (a) Fuchsian & & \multicolumn{3}{c}{(b) Quasi-Fuchsian}\\
\end{tabular}
\caption{\textbf{Simple examples of limit sets.} (a) Generators are four and mutually tangent inversion circles. The limit set is a circle; the subgroup is defined \emph{Fuchsian}. (b) If not being exactly a geometrical, but a Jordan curve, such groups are known as \emph{quasi-fuchsian}.}\label{fig_simple_examples}
\end{figure}

\label{par_invariance}Their unique nature let both families of invariant circles be gathered into the basic tool\-kit for the ex\-plo\-ra\-ti\-on of the subgroups of $\mathcal{M}$; in fact, \emph{invariance} prevents ambiguities and then it shows to be an essential property for building up mathematical theories.\footnote{For more information, see \cite[pp. 23 et ff.]{Ford-1929}.} Invariant circles are also of help to obtain graphical representations of generators (fig. \ref{fig_simple_examples}/a and \ref{fig_circle_inversion}). We will deal here with subgroups of \emph{self-inverse} and of \emph{non-self-inverse} Möbius transformations in form (\ref{eq_03}) or not respectively.

The renderings on the left and at the center inside the strip of figures (\ref{fig_simple_examples}) are known as \emph{tessellation} and \emph{limit set}, and technically obtained by computing all the chains/orbits (\ref{eq_02}) at p. \pageref{eq_02} up to some given bounded depth, with the distinctive approach to rendering all the images of the starting invariant disc through each orbit or just the last in line. \emph{Kleinian} is the definition for groups $\mathcal{G}$ of LFT (\ref{eq_01}).

The research on subgroups $\mathcal{G}$ came up in the second half of the XIX century. The initial development of this theory involved the classification rules of subgroups, according to the following properties:\mylittlelinebreak

($a$) \emph{shape} and \emph{topology} of limit sets: a circle, a general curve or a dust of points for subgroups of \emph{Fuchsian}, \emph{quasi-Fuchsian} and of \emph{Schottky} kind respectively (fig. \ref{fig_simple_examples});

($b$) \emph{numerical nature} of the coefficients $a$, $b$, $c$ and $d$; for example, the subgroup is \emph{modular} if they are real integers, or defined \emph{Picard} groups, after \'Emile Picard (1856-1941) groups, if coefficients are Gaussian integers;

($c$) special \emph{algebraic relations} between coefficients where the above mentioned trace plays as a key tool.

The shapes of limit sets $\lambda$ for subgroups of $\mathcal{M}$ are generally ruled by fractal patterns.\footnote{Like it happens to \emph{Julia sets} $\mathcal{J}$, the limits for the iteration of non-linear functions in one complex variable. These are two well-known kindred theories, based on orbits built according to the criteria of the algebraic structures which functions belong to: groups or singletons.}

\begin{figure}[!h]
\centering
\begin{tabular}{cp{0.2cm}cp{0.2cm}c}
    \includegraphics[height=3.0cm]{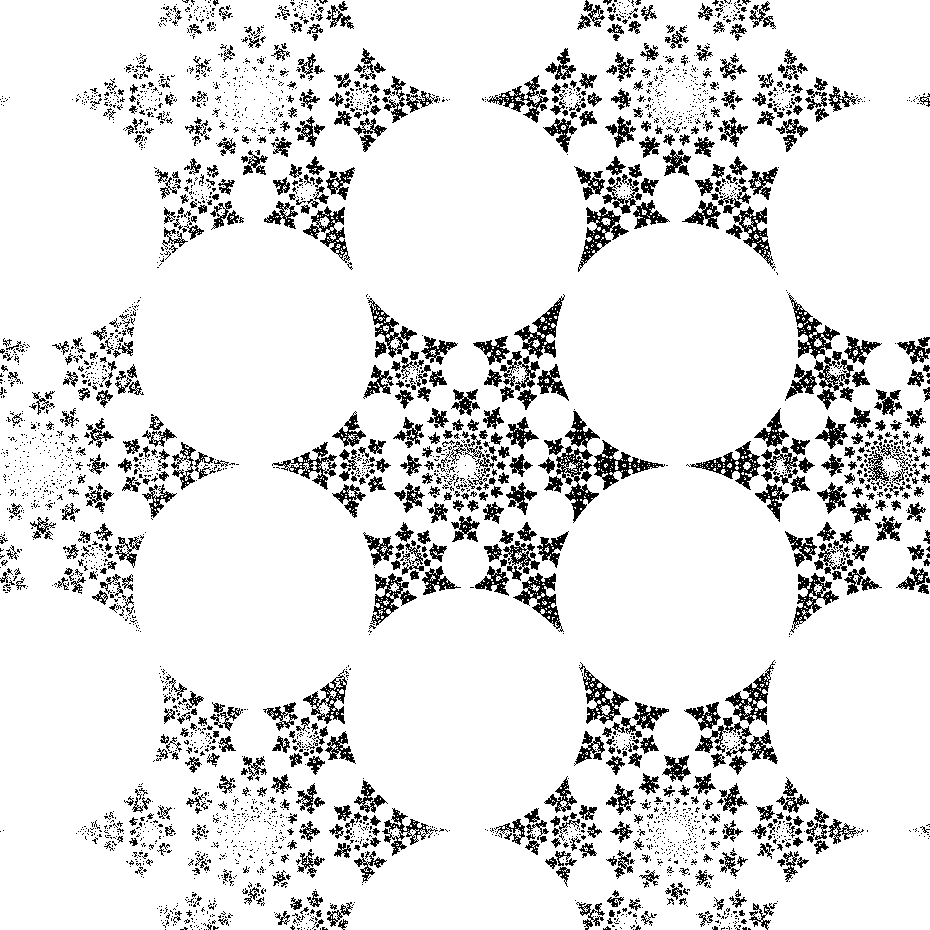}
    & &
    \includegraphics[height=3.0cm]{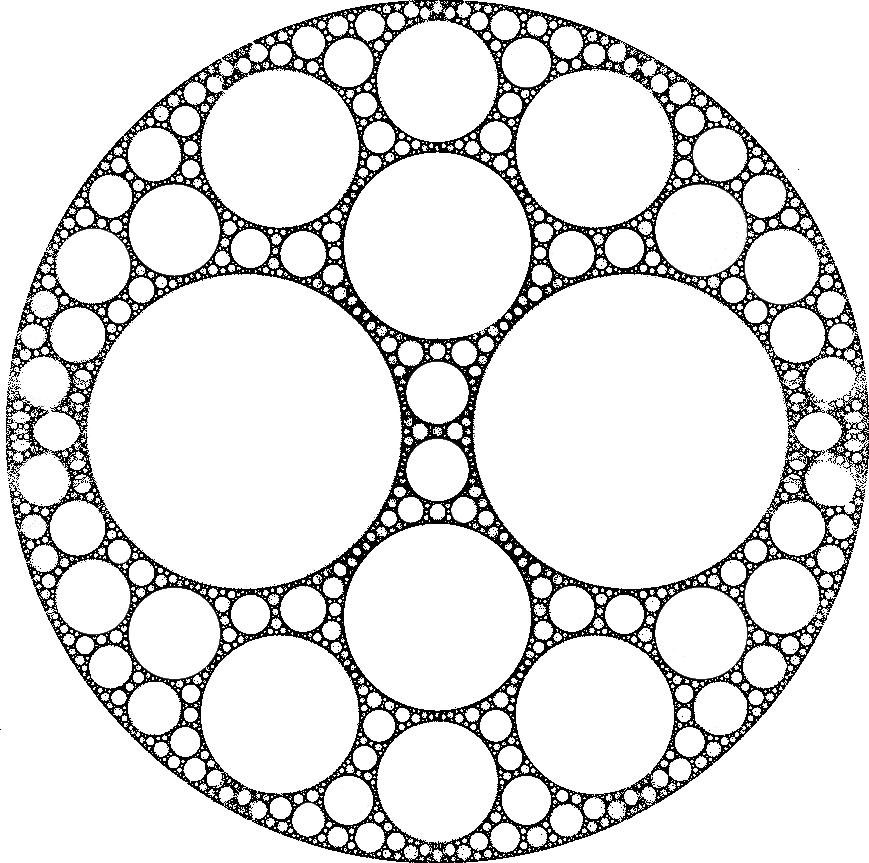}
    & &
    \includegraphics[height=3.0cm]{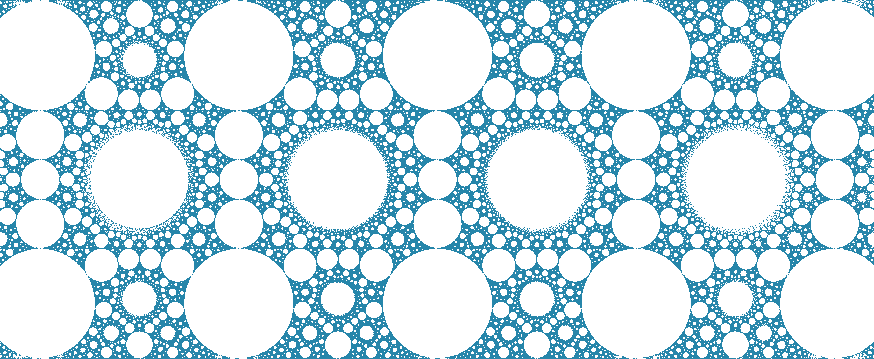}
\end{tabular}
\caption{\textbf{Circle packings.} Many limit sets for subgroups of Möbius maps spread their point around circles, which tend to pack bounded surfaces, like disks, or infinite strips.}\label{fig_circle_packing}
\end{figure}

\begin{wrapfigure}{i}{0.34\textwidth}
\centering
\includegraphics[width=0.3\textwidth]{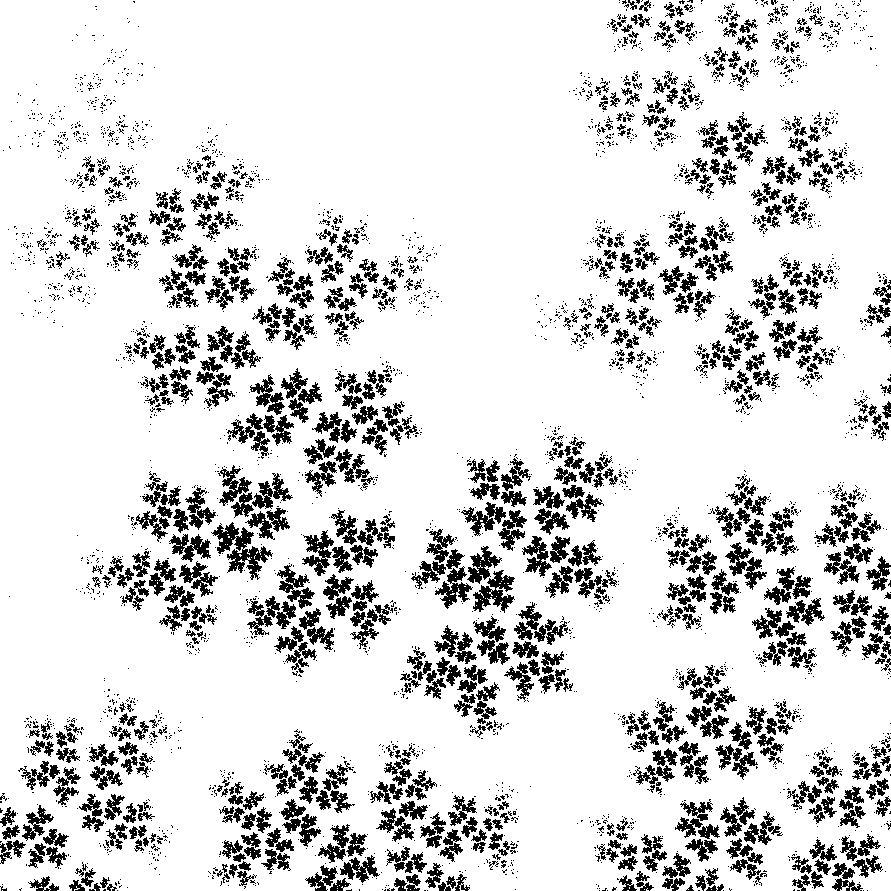} \caption{This is a close-up of the limit set for a degenerate subgroup of 2-generators, whose coefficients satisfy special and delicate numerical conditions.}\label{fig_kl_pix_02}
\end{wrapfigure}

There exist groups, defined \emph{degenerate}, which escape this pattern (fig. \ref{fig_kl_pix_02} at p. \pageref{fig_kl_pix_02}). Other groups have limit sets which spread in ways that their complement (the set of discontinuity $\Omega$) consist of circles that are said to \emph{tessellate} a given portion of the complex plane $\CC$: i.e., they \emph{cover}, or \emph{fill}, the space through a well-ordered geometric distribution. Tessellation is part of the so-called \emph{circle packing}, a collection of studies focusing on optimal\footnote{Aiming at reducing the gap between the original area and the packing to 0.} patterns for filling in areas by means of circles (fig. \ref{fig_circle_packing}). Packing could be assumed as an algorithm for approximation. We are interested here in deploying an efficient algorithm to render,  on a computer screen, the limit sets of Kleinian groups that are not \emph{elementary}, i.e., which include at least three points.

\begin{figure}[!b]
\centering
\begin{tabular}{cp{1cm}c}
    \includegraphics[height=5cm]{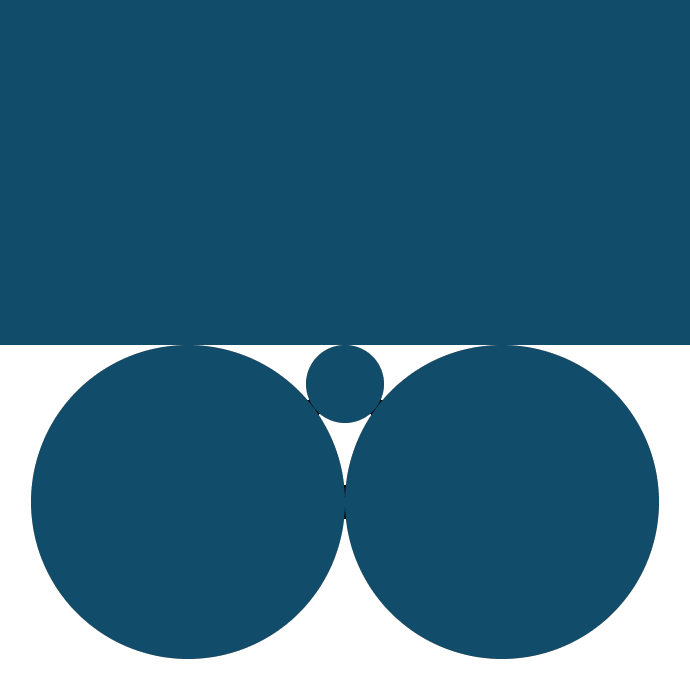}
    & &
    \includegraphics[height=5cm]{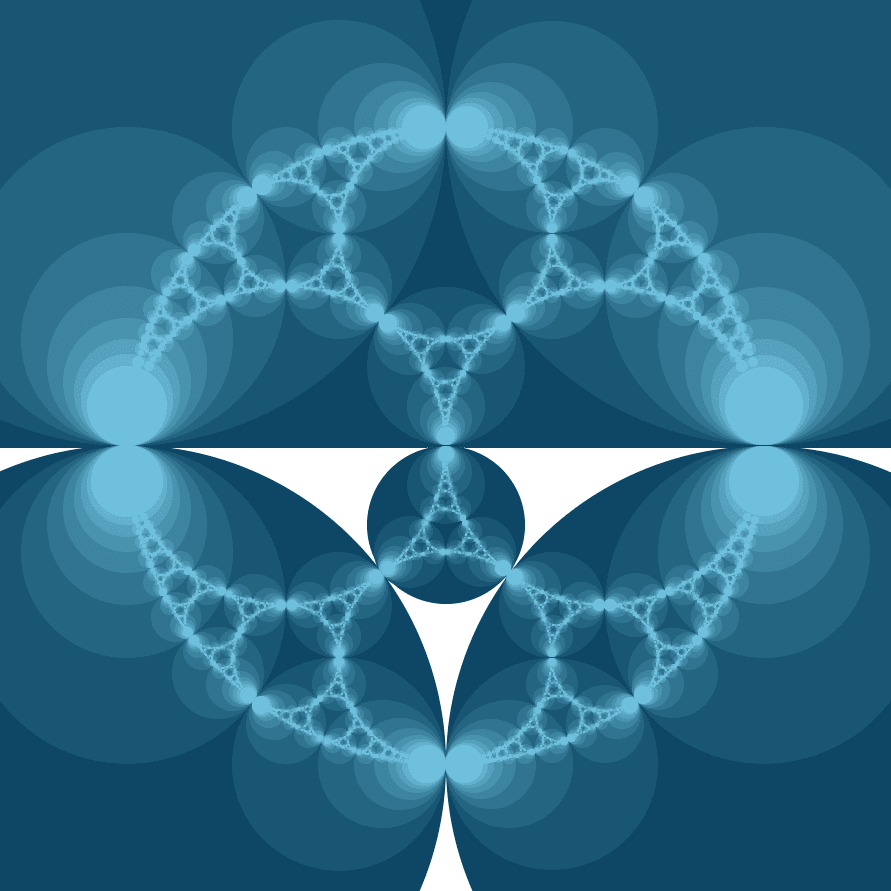}\\
    (a) Generators & & (b) Final rendering
\end{tabular}
\caption{{\textbf{Tessellation via circle inversion.} (a) A different disposition of four mutually tangent circles than fig. \ref{fig_simple_examples}/a was adopted here to work inversion maps (\ref{eq_03}). The tangency condition is preserved along the construction and it prevents images circles from overlapping each other. The gradient of lighter shades enhances convergence as well as position and shape of the limit set.}}\label{fig_discs_apollonian}
\end{figure}

\subsection{Two kinds of rendering}
The circle representation of a generator could extend to that of the chain (\ref{eq_02}) because of having the same algebraic nature as of (\ref{eq_01}), according to the discussion at \S \ref{section_env}. Would it be always worth anyway?

The renderings shown throughout this article are joined by the same goal: displaying the limit sets for subgroups $G\subset\mathcal{M}$. They show up in two kinds, and each for a precise purpose: when discs are drawn (or painted), we also chose to pick up colors out from a palette of shades being sorted by a gradient, in order to obtain a chromatic analogy for the decreasing sequence of disc sizes, for instance. Alternatively, discs are not drawn if we need to display the end points of the orbits and we are not interested in representing the whole orbit, but just (the approximation of) their final fate, the above mentioned $\Lambda_d$ of the limit set. So these two kinds of rendering are intended to highlight the \emph{behavior} and \emph{fate} of the orbits respectively. The limit set will reveal as the consequence from the decreasing size of these discs during the generation of isometric circles.

\begin{figure}[!h]
\centering
\begin{tabular}{cp{0.1cm}cp{0.1cm}c}
    \includegraphics[height=4cm]{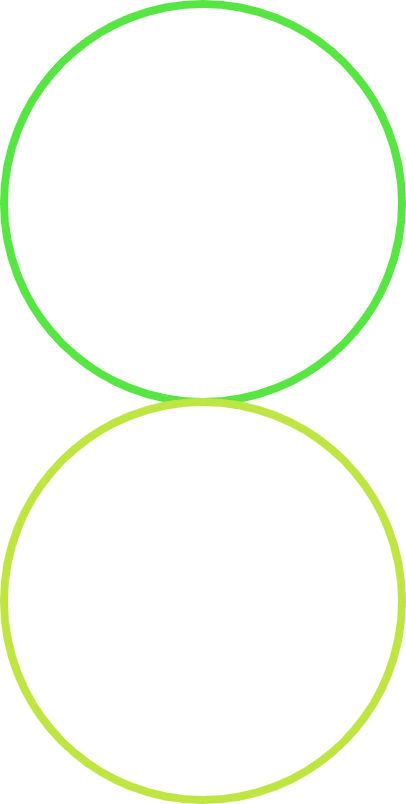}
    & &
    \includegraphics[height=4cm]{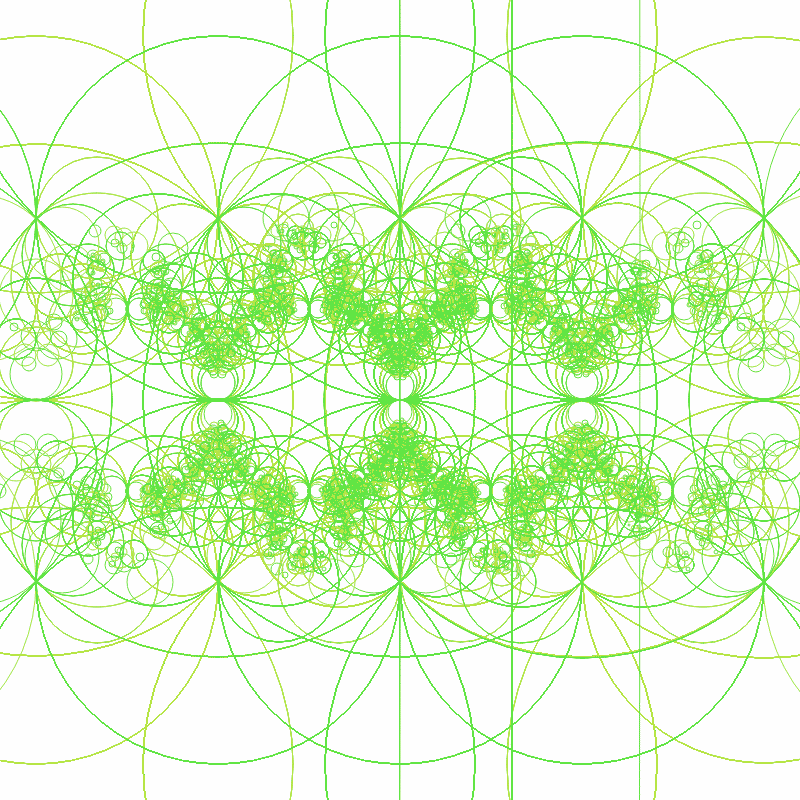}
    & &
    \includegraphics[height=4cm]{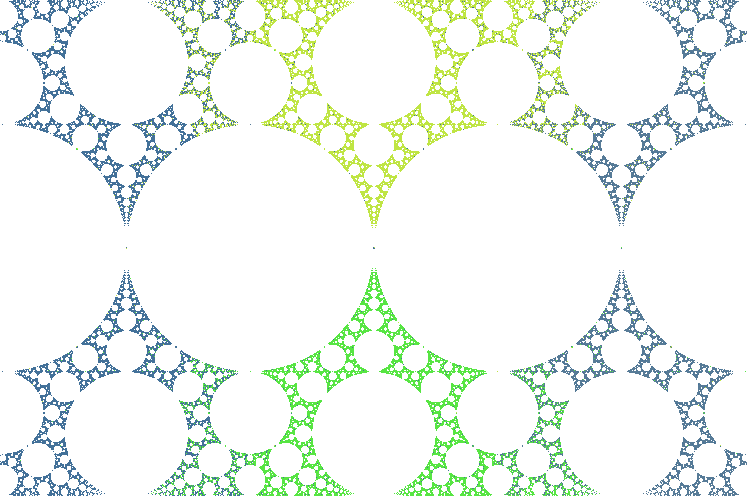}\\
    (a) Generators & & (b) Disc images & & (c) Limit set\\
    {\footnotesize isometric circles} & & {\footnotesize isometric circles} & & {\footnotesize points}\\
\end{tabular}
\caption{\textbf{Choice of proper strategy.} (a) Generators are parabolic. The subgroup action has been rendered through (b) circles (the limit set is barely recognizable) and (c) points/pixels. Colors in (c) are associated to the starting ge\-ne\-ra\-tor of each orbit.}\label{fig_riley_slice}
\end{figure}

The main reason behind the choice of rendering tesselations or limit set relies in the possibility of producing pictures that will not look as messy and confusing (fig. \ref{fig_riley_slice}/b); this last event often happens when the disc images overlap each other. 
\section{Introduction to the lexicographic approach}\label{section_lexicographic_approach}
We need an optimal computation strategy because groups processing needs to skip all the chains including contiguous pairs $g_k\circ g^{-1}_k$ of inverse generators for example, which resolve into the identity map $I(z)$ and give rise to duplicate orbits that are represented by equivalent but formally shorter chains.

The earliest renderings, via tessellations or limit sets, were already available on paper at the end of the XIX century, in the masterpiece by Fricke and Klein \cite{FrickeKlein-1897}. During the modern times of digital computing, this problem was tackled through a lexicographic approach based upon a \emph{finite state automata} (see \cite{Epstein-1992}) that generates all the chains by \emph{permutations}, each uniquely binding to one of the orbits up to the bounded maximal length/depth $l=d<\infty$. We remark that this approach was not originally conceived for computational goals, as dating back to a time when there was no enough familiarity and confidence with these problems, hence \emph{its features were not geared to optimizing speed, efficiency, and to saving memory resources}.

\begin{figure}[!h]
\centering
\begin{tabular}{cp{1cm}c}
\includegraphics[height=3.5cm]{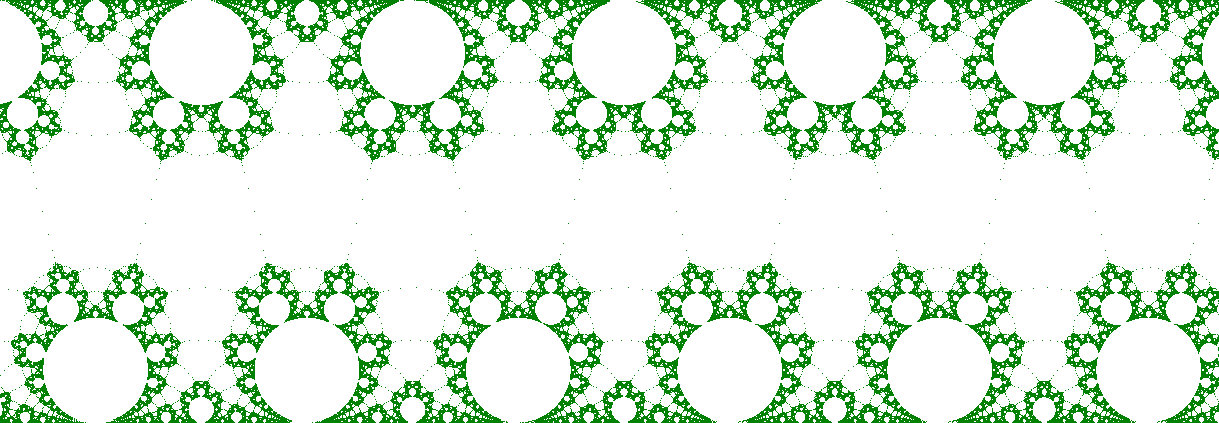}
& &
\includegraphics[height=3.5cm]{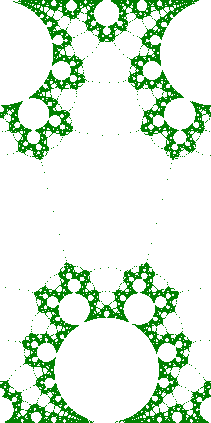}
\end{tabular}
\caption{\textbf{Tessellation and pattern.}}
\end{figure}

\subsection{Trees of words and duplicates}
\begin{wrapfigure}{i}{0.34\textwidth}
\centering
\input{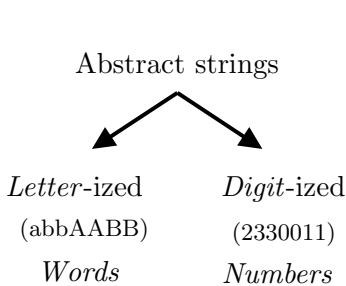}
\caption{\textbf{Materialization into trees.}}\label{fig_abstract_repr}
\end{wrapfigure}
\emph{Concatenation} is a general operation for producing \emph{abstract strings} (fig. \ref{fig_abstract_repr}): sequences of symbols, each being decoupled from semantic meaning. In mathematics, it finds to be useful for representing multiple application of functions, thus it can be extended to working with chains (\ref{eq_02}).

It underlies the so-called \emph{lexicographic} approach to the rendering of subgroups $\mathcal{G}\subset\mathcal{M}$, where strings are generated in the terms set out by the chains (\ref{eq_02}). Symbols are here assumed to be the letters $l_n$ of the Western alphabet ($a$, $b$, $c$, $d$, \dots), thus abstract strings materialize into sequences of alphabetical letters, that are known as \emph{words}. Namely, each generator of subgroups $\mathcal{G}$ will be associated to one letter as follows:

\begin{equation}\label{eq_binding_inversions}
g_1=a,\hspace{0.5cm}g_2=b,\hspace{0.5cm}g^{-1}_1=A,\hspace{0.5cm}g^{-1}_2=B;
\end{equation}

This writing lists the bindings required for applying the lexicographic approach to so-called 2-\emph{generators subgroups}.\footnote{Un\-less other\-wi\-se spe\-ci\-fi\-ed, the ex\-pres\-si\-on `$n$-ge\-ne\-ra\-tors subgroup' does not count the in\-ver\-se maps $g_n^{-1}$.} Here we first notice that letters stand out as a good choice for setting up a one-to-one relation between one and distinct symbol and one only generator in order to reproduce the inversion relationship between pairs of generators: mutually inverse generators are analogously represented by the duality of the small (lowercase) and the ca\-pi\-tal (uppercase) representation of the same al\-pha\-be\-tic symbol. Anyway, for following the next arguments, we have to remark that \emph{every such binding is just a resort and one of the possible viable choices}.

In more details, the lexicographic approach works upon a set $\mathcal{W}_n$ of $n<\infty$ symbols, each being conventionally associated to one and only one ge\-ne\-ra\-tor in the given subgroup $G\subset\mathcal{M}$. Analogously, $\mathcal{W}_n$ is termed \emph{alphabet}, as it serves to encode the chains (\ref{eq_02}) into the readable representation of a string of letters, which is contextually defined as \emph{word} and which could be read from left to right (LR) or from right to left (RL). Throughout the present article, words will be conventionally read in RL order. It is easy to check out these biunivocal connections
$$\textnormal{orbits}\hspace{0.5cm}\leftrightarrows\hspace{0.5cm}\textnormal{chains} \hspace{0.5cm}\leftrightarrows\hspace{0.5cm}\textnormal{words}.$$

\begin{figure}[!t]
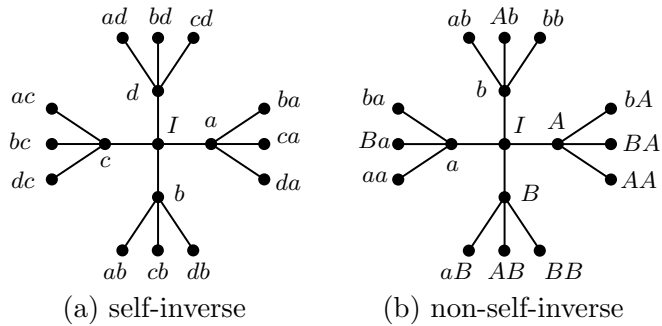

\centering
\begin{tabular}{cc}
    \input{figs/pic/tree_full_inversion_lex.pic}
    &
    \input{figs/pic/tree_full_composition_lex.pic}\\
    (a) self-inverse & (b) non-self-inverse
\end{tabular}
\caption{\textbf{Multi-branched trees up to depth 2.} Lexicographic representation of the two initial steps in the growth for the tree models associated to 2-generators groups of (a) circles inversions or (b) not.}\label{fig_trees_01}
\end{figure}

The lexicographic approach will build up a \emph{dictionary}, that is, a collection of finite length words $w$. Dictionaries are filled in by all the words being generated by appending single nodes up to some bounded length $l$.

\begin{wrapfigure}[12]{o}{0.38\textwidth}
\centering
\input{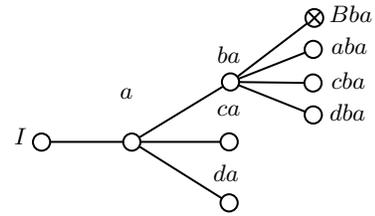}
\caption{\textbf{Tree growth with cancellations.} The identities $aA$, $Aa$, $bB$, $Bb$ stop the tree growth as they would refer to equivalent words.}\label{fig_partial_tree}\vspace{-0.5cm}
\end{wrapfigure}
In this environment, the concatenation of letters into words can be modelled through a $(4-1=3)$-branched tree (fig. \ref{fig_partial_tree}), where every new node gives rise to $n-1$ new branches. Orbits are represented by paths connecting the root to leaves; the \emph{depth} of an orbit thus amounts to the number of nodes traversed up to reaching a leaf.

Words here show up as $a$, $ABab$, $BBaaB$ for example. Like ordinary ones, each encodes \emph{one} meaning. The converse is not true. The existence of mutually inverse generators in the group definition opens to the possibility of building special words -- such as $aA$, $Aa$, $bB$, $Bb$ for 2-generators groups, which formally represent the identity map $I(z)=z$, which is generally obtained through the formal composition $g(z)\circ g^{-1}(z)$ (see \S \ref{section_env}). Operatively, the identity map does not alter the action of the chain and can be safely dropped: this management is formally carried out by the \emph{cancellation}, within the given word, of contiguous symbols that jointly pertain the identity map: for example, $Aaaa$, read from right to left, reduces to the shorter form $aa$, a word that was previously generated along a same process.

If a word includes no identities, it is said \emph{reduced}. Thus, we must separate form from action here: the possibility of cancellations within formal words hints at the existence of equivalent but shorter ones; or, similarly, to the existence of infinitely many and longer forms of a same reduced word. For what follows, we want to point out that \emph{the action of every chain} (or word, or orbit) \emph{is subjected to its formal representation}, for any set of symbols in use: in fact, cancellations are necessary for they involve redundant operations and computational costs if neglected.

\begin{figure}[!h]
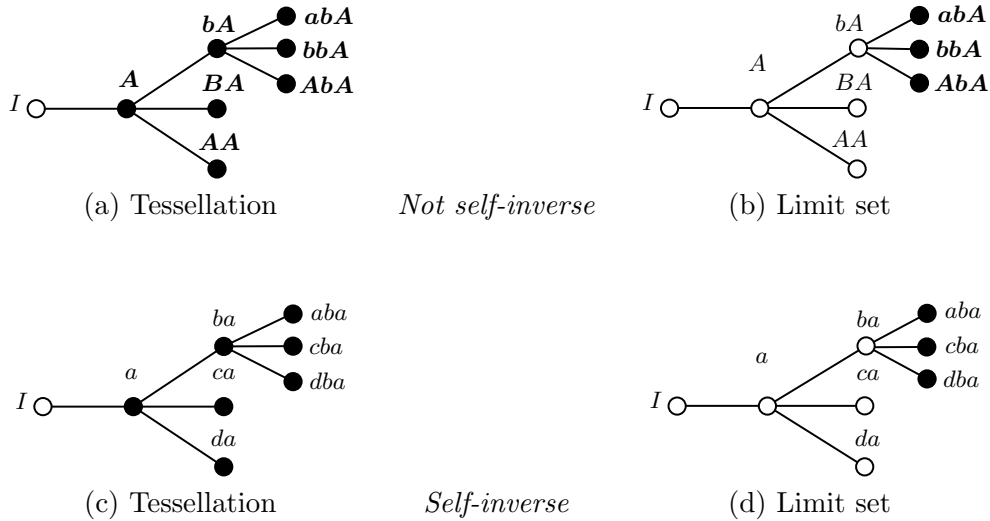

\centering
\begin{tabular}{ccc}
    \input{figs/pic/tree_partial_discs_composition.pic}
    & &
    \input{figs/pic/tree_partial_limitset_composition.pic}\\
    (a) Tessellation & \emph{Not self-inverse} & (b) Limit set\\
    \\
    \\
    \input{figs/pic/tree_partial_discs_circle_inversion.pic}
    & &
    \input{figs/pic/tree_partial_limitset_circle_inversion.pic}\\
    (c) Tessellation & \emph{Self-inverse} & (d) Limit set\\
\end{tabular}
\caption{\textbf{Partial trees up to depth 3.} The entries in bold are
those to be rendered.}\label{fig_trees_02}
\end{figure}

Given a word of given length, cancellations shall be evaluated along a cascading approach: the formal word $ababBAA$ first requires to drop $bB$ and we get $abaAA$; again, we drop $aA$ and we finally have $abA$ which includes no more cancellations; whereas we do not need this cascading check when we are building words step by step.

These remarks definitely attest that the correct processing of our subgroups requires every newly generated word to be checked for not including cancellations.

\subsection{Presentations and multiplication tables}
In terms of our tree model, this latter task calls in the concept of \emph{phyllotaxis}, i.e., the set of rules followed by the tree during its growth. Here they collect into lists that could be concise or not, depending on the degree of complication governing the tree growth. In general, rules concern how growth continues or stop. Identities are just special and simple cases of such stopping rules, all defined as cancellations for short. The expression below presents a widely used and compact form that lists generators on the left and cancellations on the right:
\begin{equation}\label{eq_04}
\langle a, b\ |\ aa=bb=I \rangle
\end{equation}
\noindent This refers to a subgroup of \emph{self-inverse} maps, such as circle inversions (\ref{eq_03}) for instance. The letter $I$, meaning to the \emph{i}dentity, is also conventionally replaced by the unit value 1, assuming the composition operator `$\circ$' to be formally read like the arithmetic multiplication. In the next example expression, we worked with subgroups (\ref{eq_binding_inversions}), where identities originate from pairs of mutually inverse generators
\begin{equation}\label{eq_05}
\langle a, b, A, B\ |\ aA=Aa=bB=Bb=1 \rangle.
\end{equation}

\noindent Rules ensure that this is a 2-generators subgroup of \emph{non-self-inverse} maps. These two examples are known as \emph{group presentations} (or \emph{defining relations}). Each identity detection rule on the right is said \emph{cancellation}, because of symbols being deleted when identities occur within a new formal word. The goal of presentations is to obtain synthetic writings for simple groups, but they turn obsolete for more complicated actions that may feature several rules for composition and cancellation of words. We recall that, according to the remark at p. \pageref{par_no_primitive_elements}, there exists no unique and irreducible presentation. For example, the word $aaBb$ resolves into the equivalent $aa$ after the deletion of the identity $Bb$, according to the rules in the table (\ref{eq_05}).

Figure \ref{fig_trees_02} shows four partial trees that are built up according to the presentations (\ref{eq_04}) and (\ref{eq_05}) respectively. All the previous examples shall not suggest that the one-to-one relation from letters to generators is the sole way to detect identities. There exists a larger casuistry overcoming the formalities of the identities discussed so far, not just implemented through the concatenation of two opposite generators: for instance, $a^3=aaa=I$.

\begin{figure}[!t]
\centering
\begin{tabular}{cp{0.4cm}cp{0.2cm}c}
\includegraphics[height=3.0cm]{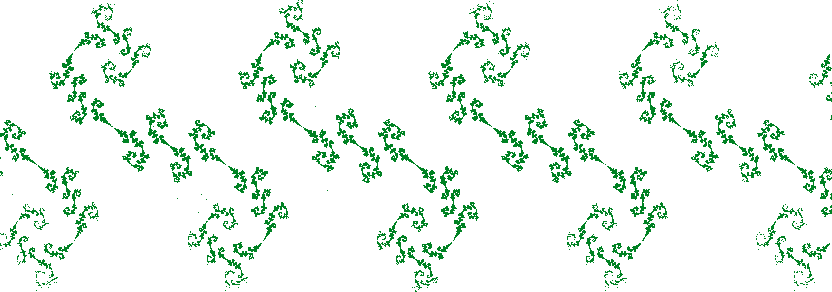}
& &
\includegraphics[height=3.0cm]{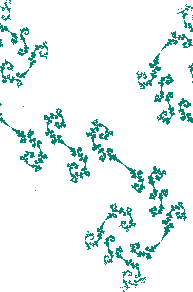}
& &
\includegraphics[height=3.0cm]{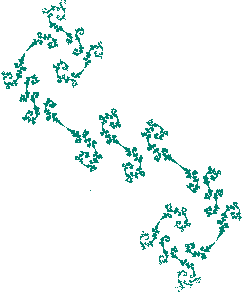}\\
(a) & & (b) & & (c)
\end{tabular}
\caption{\textbf{Tessellation and patterns.} Patterns may be wrapped into orthogonal or oblique containers.}
\end{figure}

\begin{table}[!t]
\centering
    \begin{tabular}{|p{0.2cm}|p{0.2cm}|p{0.2cm}|p{0.2cm}|p{0.2cm}|}
       \hline
       \cellcolor{gray9}
       &
       \cellcolor{gray8}
       &
       \cellcolor{gray7}
       &
       \cellcolor{gray6}
       &
       \cellcolor{gray5}\\
       \hline
    \end{tabular}\vspace*{0.2cm}

\begin{tabular}{cp{0.1cm}c}
    \begin{tabular}{|c|c|c|c|c|}
       \hline
         & \textbf{a}    & \textbf{b} & \textbf{A} & \textbf{B}\\
       \hline
       \cellcolor{gray9}\textbf{a} & a & \cellcolor{gray8}b & \dag & B\\
       \hline
       \cellcolor{gray8}\textbf{b} & a & b & \cellcolor{gray7}A & \dag\\
       \hline
       \cellcolor{gray7}\textbf{A} & \dag & b & A & \cellcolor{gray6}B\\
       \hline
       \cellcolor{gray6}\textbf{B} & a & \dag & A & \cellcolor{gray5}B\\
       \hline
    \end{tabular}

    & &

    \begin{tabular}{|c|c|c|c|c|}
       \hline
         & \textbf{a}    & \textbf{b} & \textbf{c} & \textbf{d}\\
       \hline
       \cellcolor{gray8}\textbf{a} & \dag & \cellcolor{gray7}b & c & d\\
       \hline
       \cellcolor{gray7}\textbf{b} & a & \dag & c & \cellcolor{gray6}d\\
       \hline
       \cellcolor{gray9}\textbf{c} & \cellcolor{gray8}a & b & \dag & d\\
       \hline
       \cellcolor{gray6}\textbf{d} & a & b & \cellcolor{gray5}c & \dag\\
       \hline
    \end{tabular}\\\\
    (a) $abABB$ & & (b) $cabdc$\\
    Not self-inverse subgroup & & Self-inverse subgroup
\end{tabular}
\caption{\textbf{Trasversing the multiplication table.} The tree growth in figs. \ref{fig_trees_02} and \ref{fig_trees_01} is driven by Cayley tables. The cancellations `$\dagger$' appear in the presentations (\ref{eq_05}) and (\ref{eq_04}). We can follow the zig-zag path through the upper bar with the gray shades gradient.}\label{table_mult_table_test}
\end{table}

In such a more variegated scenario, it may happen that cancellation rules could be so many to no longer  fit the goals behind the compact form of group presentations. This problem is settled by the so-called \emph{multiplication table} or \emph{Cayley table}, named after Arthur Cayley (1821--1895), which every single step along the formation of new chains of generators. Cayley tables include the same number of columns as of the generators (including the inverse ones) in the subgroup, whereas rows list \emph{all unique combinations} allowed in a given group, including the cancellations. Again, every row is announced by one combination of generators on the far left column, and it is accessed by means of the combination with the cells in the other columns, following a sort of zig-zag path eventually ending at cancellation (box \ref{table_mult_table_test}).\footnote{Presentations can be seen as \emph{synthetic} versions of the \emph{analytic} multiplication tables \cite[p. 88]{LynSch-2001}, which provide specific composition rules besides cancellations.}

The two examples in box \ref{table_mult_table_test} stand out as the simplest ever, because the only directive to follow, for trasversing the table, wants to replace the current symbol by the next in line, again and again up to the chosen maximal length of the orbit or until we do not stumble into a cancellation rule. The (RL) reading of the word $abA$ is equivalent to the following path $A\underset{b}{\rightarrow}b\underset{a}{\rightarrow}a$, running through three rows, one per each symbol. There exist groups whose multiplication tables include rows that are announced by words being longer than one symbol (see the example at \cite[p. 359]{Indras-2002}), such as the path $a\underset{b}{\rightarrow}ba\underset{A}{\rightarrow}Aba$ that runs over the table rows announced by $a$, $ba$, and $Aba$.\footnote{ The formal word $Bbba$ will eventually meet the cancellation rule in the row announced by the letter $B$.} This is an excerpt of a multiplication table including longer entries than one letter \cite[p. 359]{Indras-2002}:

\begin{table}[!h]
\centering\em
\begin{tabular}{|p{3cm}|c|c|c|c|}
\hline
& a & b & A & B\\
\hline
bAB & I & I & BA & B\\
\hline
Bab & ba & b & I & I\\
\hline
\end{tabular}
\end{table}

In any version, either as presentations or as tables, the tests for reduced words validation by means of cancellations cannot be exempted from implementation because of being strictly required to ensure the correct processing; otherwise said, ignoring the cancellation tests would bring inaccurate results. \noindent \emph{The lexicographic approach \textnormal{pre-processes} words: it builds them in progression and checks if the newly appended symbol has met an identity rule and then triggered a cancellation}.

During the early 1990s, digital pictures of limit sets were produced through the lexicographic approach in a few works, such as in Manna and Vicsek's \cite{Manna-1991}, Bullets and Mantica's \cite{BullMan-1992}, McShane, Parker and Redfern's \cite{McPaRe-1994}, and Parker's \cite{Parker-1995}. None of them hit the technical details of the rendering anyway. This gap in the literature was filled inside the book \emph{Indra's pearls} by Mumford, Series and Wright \cite{Indras-2002}, published in 2002 and providing a very extensive and plain discussion of the lexicographic approach. 
\section{Drawbacks of the lexicographic approach}\label{section_drawbacks} Given $n$ generators and chains (\ref{eq_02}) of maximal depth $d$, tessellations and limit sets require $\displaystyle N_T=\sum_{i=1}^d n^i$ (intermediate nodes and leaves) and $N_L=n^d$ words (leaves only, i.e. the number of permutations) respectively. The lexicographic approach features the following additional computation costs:\mylittlelinebreak

(1) a \emph{table} for binding letters to the indexes of the generators stored in an array;

(2) a \emph{table} for registering the association between the letters of generators and of their inverses;

(3) the \emph{implementation} of bread-first (equivalently, depth-first) \emph{algorithm} \emph{for trasversing the tree of words} for generating the words dictionary up to a finite length;

(4) the \emph{memory space} required to store the dictionary;

(5) the \emph{translation of symbols into indexes} to pick up each generator long the chain of compositions.\mylittlelinebreak

\noindent The following tables report the memory sizes of dictionaries including words up to length 17, which could allow some rendering quality. And these costs are doomed to dramatically grow, especially if close-ups of the limit set have to be rendered. Hence compiling the dictionary would equivalently turn into a very expensive process, that demands long computation times and very huge memory resources.

\begin{table}[!h]
\centering\footnotesize
    \begin{tabular}{|l|l|l|l|l|l|l|l|l|l|l|}
    \hline
    \emph{words length} & 0 & 1 & 3 & 5 & 7 & 9 & 11 & 13 & 15 & 17\\
    \hline
    \multicolumn{10}{l}{\textbf{}}\\
    \multicolumn{10}{l}{\textbf{Tessellations}}\\
    \hline
    \emph{process steps} & 1 & 5 & 53 & 485 & 4373 & 39365 & 354293 & 3188645 & 28697813 & 258280325\\
    \hline
    \emph{dictionary size} & 1B & 5B & 53B & 485B & 4.2KB & 38KB & 346KB & 3.04MB & 27MB & 246.3MB\\
    \hline
    \multicolumn{10}{l}{\textbf{}}\\
    \multicolumn{10}{l}{\textbf{Limit set}}\\
    \hline
    \emph{process steps} & 1 & 4 & 9 & 81 & 729 & 6561 & 59049 & 531441 & 4782969 & 43046721\\
    \hline
    \emph{dictionary size} & 1B & 4B & 9B & 81B & 729B & 6.4KB & 57.7KB & 519KB & 4.56MB & 41.06MB\\
    \hline
    \end{tabular}
\caption{\textbf{Memory size for words}. The upper and the lower table have been compiled for groups of four generators of self-inverse maps and of non-self-inverse maps respectively.}\label{table_count_dict_discs}
\end{table}

\noindent The need of huge memory loads was already pointed out at \cite[p. 141]{Indras-2002}, where three approaches were provided to work around this problem: one was based upon recursion, the others on the tree model. All require considerable resources in terms of function calls stack. The recursion-based approach looks as the most onerous in this sense, as it trig\-gers as ma\-ny calls as the dic\-tio\-na\-ry si\-ze, i.e. $N_T$ or $N_L$; whereas the other two ap\-proa\-ches look ra\-ther com\-pli\-ca\-te and ex\-pen\-si\-ve, because discarding out the dictionary would imply the constant tracking the paths in tree while they have to be travelled back and forth in order to visit all the nodes therein. \emph{The lexicographic approach was not originally devised to saving resources so that the performance would eventually slow down during the running}. Any alternative should then aim at giving a lighter and quicker approach, i.e. in practice, at dropping orbits storage and at devising alternatives to step-by-step orbit generation. 
\section{Index generation: the numerical
alternative}\label{section_numerical_alternative}
In order to get away from the lexicographic environment, we shall step back to the abstract level of composition, relying upon the abstraction of symbols, as discussed in \S \ref{section_lexicographic_approach} (fig. \ref{fig_abstract_repr}). We recall that letters are just meant as a choice for opening to rendering operations. A different materialization of abstract symbols consists in involving digits and numbers instead. We start from reviewing the insertion of tree nodes in fig. \ref{fig_trees_01} under this new perspective.

According to the \emph{Basis Representation Theorem} \cite[pp. 8--9]{Andrews-1971}, every numerical quantity $Q$ can be written as a unique string $q_b$ in base $b\geq 2$. $Q$ acts like an abstract concept that allows travelling through arbitrary representations. $Q$ is invariant under base conversion, and only the formal appearance changes; thus the increasing (or decreasing) trend of sequences of numbers in base 10, say $1_{10}$, $2_{10}$, $3_{10}$, \dots, will be kept up the same trend under another numerical base. Given $q_2$ and $q_{10}$, then $q_{10}\rightarrow Q\rightarrow q_2$: for every integer in base 10, there exists one and only one conversion into a different base. Base conversion represents an \emph{unambiguous and reliable} approach to the formalization of orbits/chains.

\begin{table}[!h]
\centering
\begin{tabular}{cccccccccccccccc}
$0_{10}$ & $1_{10}$ & $2_{10}$ & $3_{10}$ & $4_{10}$ & $5_{10}$ & $6_{10}$ & $7_{10}$ & $8_{10}$ & $9_{10}$ & $10_{10}$ & $11_{10}$ & $12_{10}$ & $13_{10}$ & $14_{10}$ & $15_{10}$\\
\vspace{0.2cm}
$0_4$ & $1_4$ & $2_4$ & $3_4$ & $10_4$ & $11_4$ & $12_4$ & $13_4$ & $20_4$ & $21_4$ & $22_4$ & $23_4$ & $30_4$ & $31_4$ & $32_4$ & $33_4$\\
\end{tabular}\vspace{-0.2cm}
\caption{\textbf{Conversion from base 10 to 4.}}\label{table_base_conversion_01}
\end{table}

We notice that every base $n$ is equipped with a set of exactly $n$ distinct digits, which we could, even if improperly, call as \emph{alphabet} again, because of performing homologous tasks.

\begin{table}[!h]
\centering
    \begin{tabular}{lp{0.6cm}p{0.6cm}p{0.6cm}p{0.6cm}|cccc}
        \emph{Generator} & $g_1$ & $g_2$ & $g^{-1}_1$ & $g^{-1}_2$\\
        \emph{Symbol} & a & b & A & B\\
        \emph{Array index} & 0 & 1 & 2 & 3\\
        \emph{Cancellations} in letters & aA & bB & Aa & Bb & aa & bb & AA & BB\\
        \emph{Cancellations} in digits & 02 & 13 & 20 & 31 & 00 & 11 & 22 & 33\\
        & \multicolumn{4}{c}{\emph{non-self-inverse} generators} & \multicolumn{4}{c}{\emph{self-inverse} generators}
    \end{tabular}
\caption{\textbf{Environmental arrays for 4-generators groups.}}\label{table_env_arrays}
\end{table}

The transition from the lexicographic to the indexed approach has been depicted in table \ref{table_env_arrays}, by comparing formal compositions.

\begin{table}[!b]
\footnotesize

\begin{tabular}{p{0.36\textwidth}|p{0.58\textwidth}}
    \begin{tabular}{p{1.1cm}p{0.7cm}p{0.7cm}p{0.7cm}p{0.7cm}}
    Level 0 & I &  &  &\\
    Level 1 & A & B & a & b\\
            & $0_4$ & $1_4$ & $2_4$ & $3_4$\\
    Level 2 & AA & AB & Ab\\
            & $00_4$ & $01_4$ & $03_4$\\
            & BB & BA & Ba\\
            & $11_4$ & $10_4$ & $12_4$\\
            & aa & aB & ab\\
            & $22_4$ & $21_4$ & $23_4$\\
            & bb & bA & ba\\
            & $33_4$ & $30_4$ & $32_4$
    \end{tabular}
&
    \begin{tabular}{p{1.1cm}p{1.65cm}p{1.65cm}p{1.65cm}p{1.65cm}}
    Level 0 & I &  &  &\\
    Level 1 & a & b & c & d\\
            & $0_4=0_{10}$ & $1_4=1_{10}$ & $2_4=2_{10}$ & $3_4=3_{10}$\\
    Level 2 & ab & ac & ad\\
            & $01_4=1_{10}$ & $02_4=2_{10}$ & $03_4=3_{10}$\\
            & ba & bc & bd\\
            & $10_4=4_{10}$ & $12_4=6_{10}$ & $13_4=7_{10}$\\
            & ca & cb & cd\\
            & $20_4=8_{10}$ & $21_4=9_{10}$ & $23_4=11_{10}$\\
            & da & db & dc\\
            & $30_4=12_{10}$ & $31_4=13_{10}$ & $32_4=14_{10}$
    \end{tabular}\\
\centering self-inverse generators & \centering non-self-inverse generators
\end{tabular}
\caption{\textbf{Applications to subgroups}. Some entries have been skipped because of cancellation rules. Index generation covers all combinations yielded by the lexicographic approach (fig. \ref{fig_trees_01}).}\label{table_conversion}
\end{table}

At the initial stage, the difference regards the adopted symbols only. Every new number in base 4, for example, shows up as the concatenation of digits, taken from the set [0, 1, 2, 3] (table \ref{table_base_conversion_01}). In every 4-generators subgroup, the letters $a$, $b$, $c$, $d$ are respectively associated to the digits of the 4-base number system: 0, 1, 2, 3.
The formal expressions of numerical quantities under some given value are permutations of digits. Let the value 10000, then all smaller numbers from 0000 to 9999 are permutations of all digits in the 10-base system. Given an alphabet of cardinality $n=4$, the concatenation of letters up to depth $d$ is equivalent to writing all numbers in base $n$ up to the value $n^d$, in other terms, to compute all the \emph{permutations} of $n$ symbols up to depth $d$.

\emph{The generation of all words in the lexicographic approach can be completely replaced by the increasing sequence of positive integers in some arbitrary base system}. We no longer need to build up dictionaries and store massive bulks of data: the $n$-base representation can return the same information as from the combinations built on purpose through the step-by-step concatenation of letters. Numbers devolve into sequences of digits, i.e. of strings being managed through the symbols concatenation; they however keep all we need to render the subgroup action.

\begin{figure}[!h]
\centering
\includegraphics[height=3.3cm]{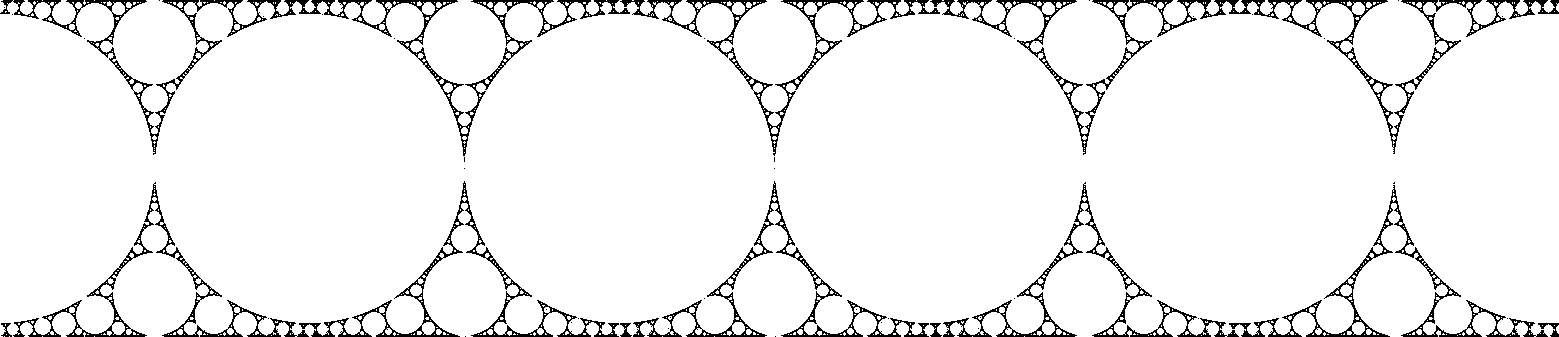}
\end{figure}

\subsection{Questioning on zero-based arrays.}
Speed would be meaningless without wise management. In fact, we are \emph{switching from quantity, represented by numbers, to quality} (i.e., the visual appearance) \emph{of digits, i.e. of symbols that are deprived from the original meaning}. This transition moves from positional numerical systems to strings of concatenated symbols. Here we could stumble into the question to working with the ambiguous role played by the digit 0 (table \ref{table_env_arrays}), which, on one side, it unequivocally refers to the action of the Möbius map stored in the \emph{array} at the index (\footnote{Arrays are data structured endowed with zero-based indexing.}) 0 but, on the other, the vanishing nature of 0 gives rise to several ambiguous but operatively equivalent formalizations: we mean to chains being prefixed by arbitrary many zeros, such as 1 for instance and all the infinitely many concatenations resumed into the periodic form $\overline{0}1$, for example. Again, \label{par_zero_filling}The digit 0, unlike all others from 1 to 9, could relate to quantification or not, depending on the position within the string of digits: it plays the multiplicative role if \emph{ap}pended to the far right (ex: 10000), or none if \emph{pre}pended to the far left (ex: 00001); we mean to \emph{trailing} and of \emph{leading zeros} respectively.

In the formal context of symbols concatenation, the digit 0 drops the role played in the positional representation of numbers; being no longer a numerical value but just as a symbol, it is an index that refers to a ge\-ne\-ra\-tor for the given subgroup; this is another reason why \emph{leading zeros are as important as trailing ones here}.\footnote{We see that the zeros left padding does not occur for groups of self-inversions for instance, where the composition of words does not allow contiguous symbols repetition.}

\begin{figure}[!b]
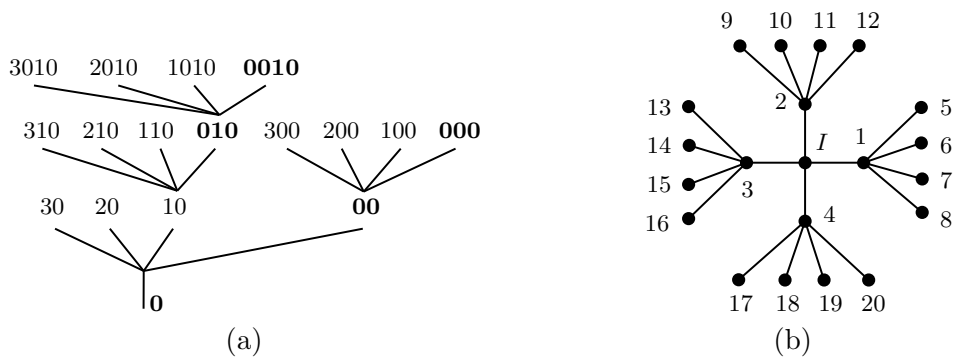

\centering
\begin{tabular}{cp{1cm}c}
        \input{figs/pic/tree_composition_indexsearch_longer.pic}
        & &
        \input{figs/pic/tree_full_numerical_base_4.pic}\\
        (a) & & (b)
\end{tabular}
\caption{\textbf{Viewpoints.} (a) \emph{Ordinal}: each integer is translated and its quantity is kept up during the translation. (b) \emph{Cardinal}: orbits are counted while they are generated; every integer is translated and each new digit in the new base representation is eventually remapped to zero-based indexing for all the integers that are associated to orbits of maximal chosen length, here 2.}\label{fig_trees_03}
\end{figure}

Would it be a real or an apparent difficulty anyway? Response is mixed and mostly affected by the way we decide to deal with zero-based indexing management of data arrays.

According to the above approach, managing the zero digit boils down to dealing with symbols concatenation, analogously to what we formerly did with letters, considering that we dealing with the family of zero-prefixed strings, like `0001', in a hybrid form where the zero digit is simultaneously worked out as a symbol in the chain, in order to compute the orbit, and as a quantity that could be prefixed by arbitrarily many zeros, as the unit, 1, is quantitatively equivalent to 01, 001, \dots\ . In short, this scenario reads every new integer, in the sequence 0, 1, 2, 3, 4, \dots and as \emph{cardinal} number and then to a numerical value that is open to multiple representations (fig. \ref{fig_trees_03}/a).\footnote{The transformation from numerical values to strings of symbols is one-to-many (= multi-valued); conversely, strings with leading zeros would be encoded back to the same numerical value in the many-to-one fashion (= single-valued). Because of the aforementioned reasons, base conversion cannot cover strings with leading zeros; hence we have to implement a separate procedure for managing these special strings.}

Otherwise, we could read the input integers as \emph{ordinal} numbers, that is, we just want to \emph{count} the orbits in the order of appearance during the process of base conversion and bind an increasing number. Hence the strictly positive integers 1, 2, 3, 4, 5, 6, \dots, will just indicate different orbits, each of which be again represented under the chosen $n$-based system,
$$1_4, 2_4, 3_4, 10_4, 11_4, 12_4, \dots$$
but regardless of the leading zeros now (fig. \ref{fig_trees_03}/b). With regard to the $4$-base representation here for instance, we need to simply re-map the indexes to the zero-based indexing as follows, in order to correctly manage arrays of digital data and pick up the Mobius maps for computations:

\begin{table}[!h]
\centering
\begin{tabular}{cccc}
1 & 2 & 3 & 4\\
$\downarrow$ & $\downarrow$ & $\downarrow$ & $\downarrow$\\
0 & 1 & 2 & 3
\end{tabular}
\end{table}

The cardinal version runs up to the 10-based integer $n^d$, where $n$ counts the symbols in the new base representation and $d$ is the maximal depth/length of the orbits, i.e., the value $n^d$ in base 10, as previously remarked, represents the set of all permutations of strings including $n$ symbols (the numerical base) and with length $d$. None of these two options represents the best choice: both are valid and the difference just regards about base representations filled by leading zeros and thus involving a larger number of orbits displayed; thus the cardinal approach will be more accurate and the ordinal one will be quicker.\mylittlelinebreak

\subsection{Guidelines for Index generation.} Resuming, the implementation of the index generation algorithm consists in\mylittlelinebreak

($1^\circ$) \emph{taking on a number} in base 10, say $93_{10}$;

($2^\circ$) \emph{encoding it} into the new base, say $1131_4$;

($3^\circ/a$) \emph{Cardinal} approach: \emph{left padding every string} yielded in the step 2 through a sequence of leading zeros up to a given finite maximal length. Suppose the latter is 8, the 4-base number obtained above would increasingly left padded in order to obtain the four strings $(8-4=4)$: $01131_4$, $001131_4$, $0001131_4$, $00001131_4$;

($3^\circ/b$) \emph{Ordinal} approach: \emph{remap every digit in the resulting base conversion} from 1-based to 0-based indexing, i.e., by decrementing each digit by one.

(4) \emph{feeding} the resulting string in the new base to \emph{the rendering engine}.\mylittlelinebreak

All process boils down to converting numbers into the new base and processing the obtained strings. We no longer need to store them. \emph{The index generation algorithm \textnormal{post-processes} words: the base conversion yield a new words which is checked whether there are subsets triggering cancellations}.

\begin{figure}[!h]
\centering
\begin{tabular}{ccc}
\includegraphics[height=5.0cm]{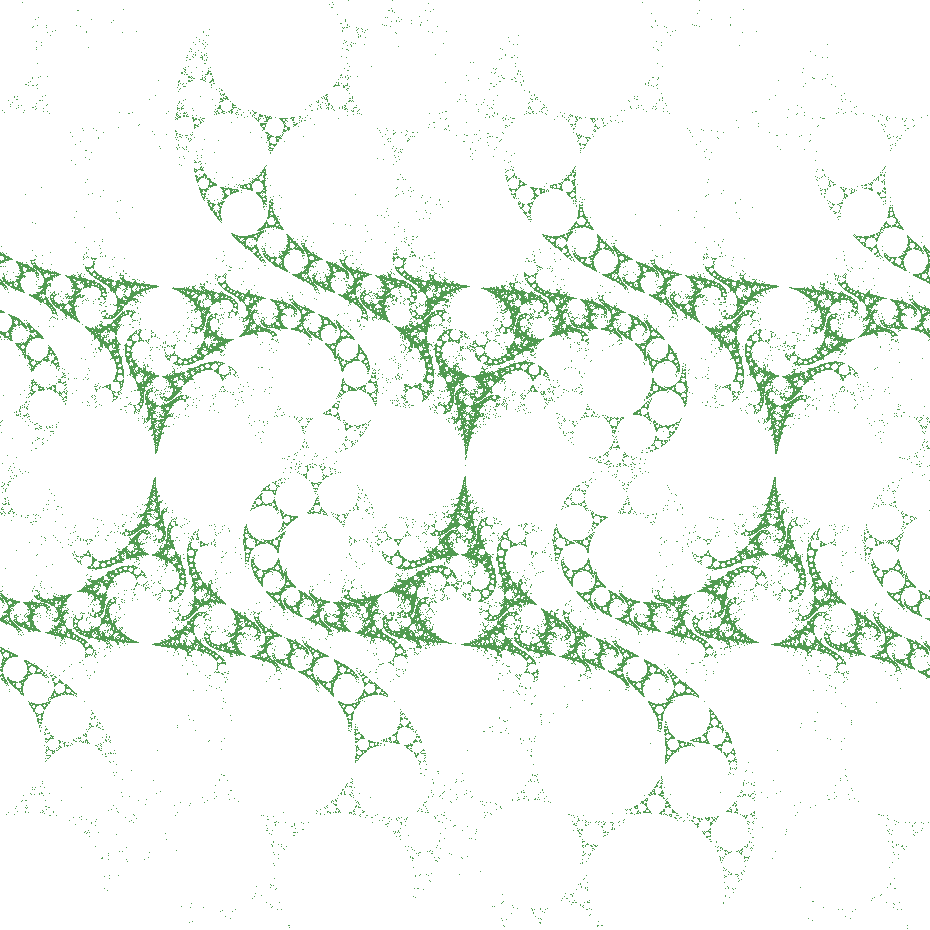}
&
\includegraphics[height=5.0cm]{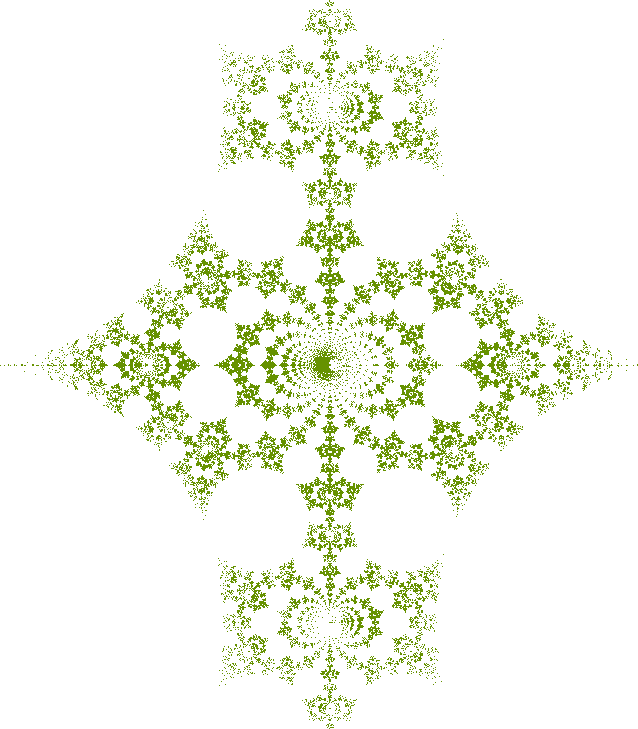}
&
\includegraphics[height=5.0cm]{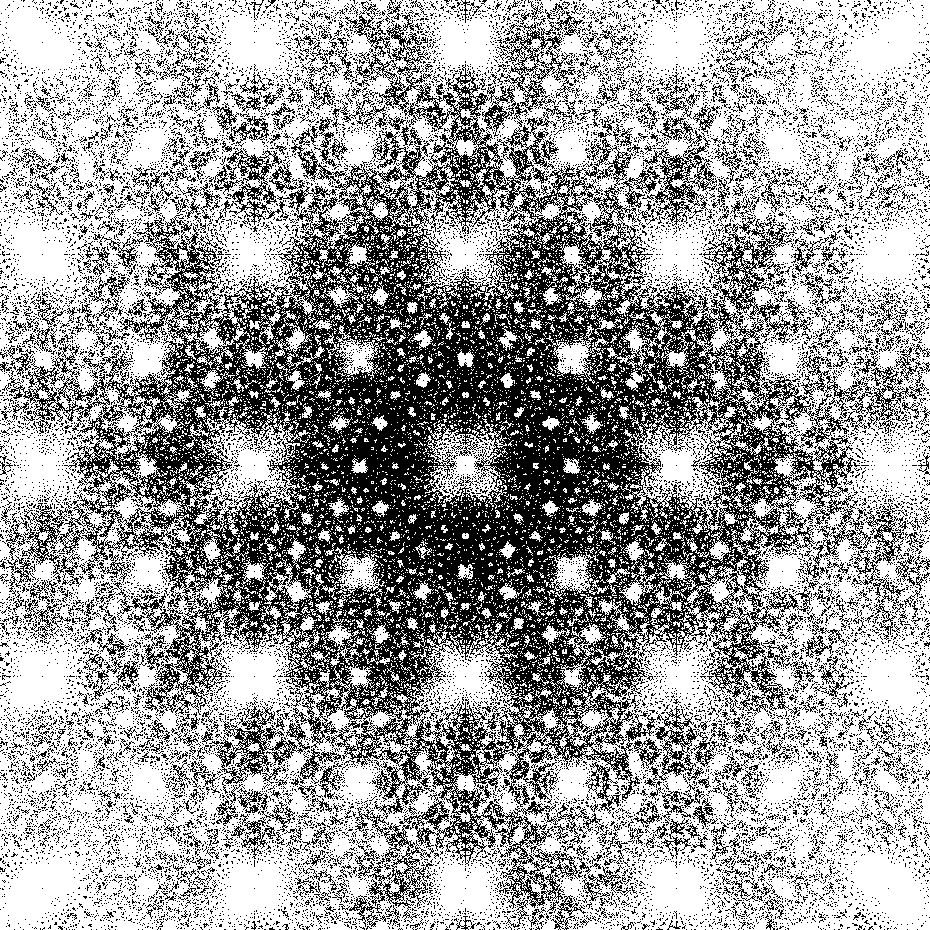}
\end{tabular}
\caption{Borders look blurred because points in the farther regions are reached in the conclusion of the process and thus may be partially covered by the relatively longer orbits. Closer points belong to regions which are more probably covered by the chosen maximal length of the orbits.}\label{fig_kl_pix_01}
\end{figure} 
\section{The pseudo-code implementation}\label{section_pseudocode}
We will give here some guidelines for the code implementation of the \emph{index generation} algorithm, in order to render tessellations or limit sets.
We split code into blocks, which might be of help for easily following every step of the process. We have adopted a pseudo object-oriented\footnote{Here closer to the syntax of C-like family of imperative languages, such as Java, Javascript, \dots.} language in order to ease the customization into the preferred environment. Generators are assumed to be instantiations of some class endowed with \emph{methods} and \emph{data containers}, accessed via this conventional syntax: \texttt{obj.\textlangle{}method\_id\textrangle{}(parameters)} for methods and for \texttt{obj.\textlangle{}variable\_id\textrangle{}} for containers (i.e., variables) respectively.

We begin from the generators listed in table \ref{table_env_arrays} at p. \pageref{table_env_arrays}: for sake of simplicity and with no loss of generalization, we will assume to work with 2-generators groups ruled by the presentation (\ref{eq_04}) or (\ref{eq_05}). The \emph{index generation} algorithm is \emph{scalable} and not affected by the cardinality of the generators set.\footnote{ In what follows, the expression \emph{index generation} refers to the algorithm, whereas the sole \emph{index} to the position within the array of generators.}

The generator--index association, within the array storage, is automatically set up by the instantiation of the logical array (table \ref{table_env_arrays} at p. \pageref{table_env_arrays}) and it is zero-based. (Hence we shall choose one of the paths discussed in the previous section: \emph{cardinal} or \emph{ordinal}.) Some environmental variables and containers have to initialized, like in the code below.

\input{text/code/0.cod}

We stress that the numerical nature of this algorithm disengages from the concept of word depth, which is more naturally tied to the transversion of the lexicographic approach. We will deal with the number of steps instead, and so we have to set up an arbitrary maximal value that stops the main loop. We opted to use the \texttt{for} syntax as it looks conceptually close to the increasing sequences of integers here involved by the base transformation.\footnote{There is no impediment against the usage of \texttt{while}-loop syntax anyway, being, as known, the generalization of the specialized conditions in the header of \texttt{for} loops.} We will work with inversion circles and with pixels/points for rendering tessellations or limit sets respectively.

\input{text/code/1.cod}

\noindent We are going to render limit sets first: a string of symbols is returned for each value of the loop counter \texttt{\_i} and processed by composition of Möbius maps.

\begin{figure}[!b]
\centering
\includegraphics[height=3.3cm]{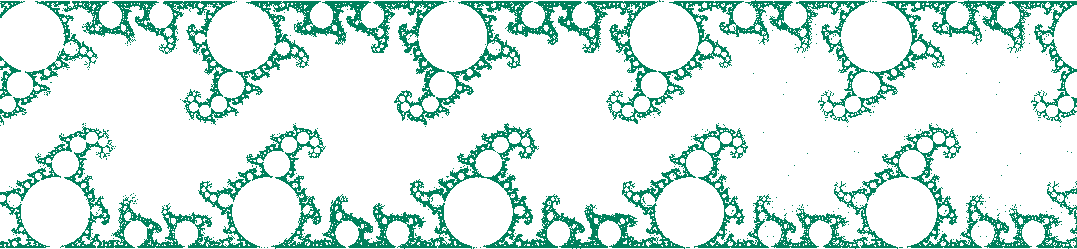}
\end{figure}

\noindent\emph{Input points.} Unlike the rasterized rendering of Julia sets, we do not need to check every point/pixel inside the region of interest (the \emph{escape time} method). According to the theory of \emph{function groups} (which both fuchsian and kleinian kinds belong to), it is sufficient to give an arbitrary input value: \emph{the definition of a limit point depends only on the sequence of elements of the subgroup $\mathcal{G}$, and not on the points belonging to the region $U$ where the action of $\mathcal{G}$ is freely discontinuous} \cite[p. 22, D.3]{Maskit-1988}. Moreover, since \emph{the limit set is transformed into itself by the subgroup transformations} \cite[p. 43]{Ford-1929}, we found worth picking up the input values from the fixed points of a generator.

The generic label \texttt{\#2.x} refers to the code block \texttt{\#2.1} which renders the \emph{limit set}. After pro\-ces\-sing the in\-put word from right to left, we will dis\-play on the screen the last element of every orbit exclusively, as required by the definition of limit sets.

\input{text/code/sub_1.cod}

\noindent\emph{Tessellation} via \emph{disc images} renderings need the reference \texttt{\#2.x} to be replaced by the block \texttt{\#2.2}, where every new inversion circle is plotted when a new symbol along the input word is read from right to left, and processed:

\input{text/code/sub_2.cod}

\noindent The label \texttt{<optional-call-to-sub-routine-\#3:leading-zeros-management>} refers to the pseudo-code implementing, in respect of the above \emph{cardinal} approach, the elaboration of strings with leading zeros too: we simply ge\-ne\-ra\-te them by pre-pending the 0 to every string from each one yielded in the main \texttt{for}-loop: for example, the input integer $\texttt{5}_{10}$ turns into $\texttt{11}_4$ in base 4 and then pad it up to maximal depth, say 6 here, by leading zeros, so to obtain: 011, 0011, 00011, 000011.

\begin{figure}[!b]
\centering
\includegraphics[height=3.3cm]{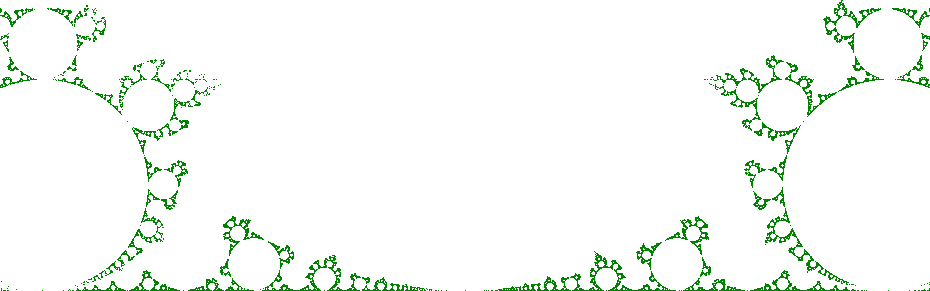}
\end{figure}

\input{text/code/3.cod}

\noindent The \emph{ordinal} version needs not to run this last subroutine. We also remark that the pseudo-code

\begin{extmdframed}
\begin{Verbatim}[fontsize=\footnotesize,frame=none]
if ( <call-to-sub-routine-#1.x:cancellation-rule-test_of_proc_str> ) continue;
\end{Verbatim}
\end{extmdframed}

\noindent refers to tests to be performed according to the Cayley table or presentation related to the given subgroup. Two examples (about the tables presented in the box \ref{table_mult_table_test} at p. \pageref{table_mult_table_test}) follow below. The equivalent cancellation rules are (\ref{eq_04}) and (\ref{eq_05}) in terms of presentations. Implementation is easy: instead of trasversing rows and columns in the table, these cancellation tests check whether every new input string of digits includes at least one of the rules on the right of the presentation.

\begin{extmdframed}
\begin{Verbatim}[fontsize=\footnotesize,frame=none,numbers=left]
//<sub-routine-#1.1:group-presentation>
function __check__group_presentation__( _digitized_word = "" )
{
<let a boolean flag and set it to 0>

<for each entry inside the group presentation>
    <check if the input digitized word includes the current entry>
    <if so, set the above flag to 1 and break this loop>
<end-of-for-loop>
        
<return the boolean flag>
}
\end{Verbatim}
\end{extmdframed}

\begin{figure}[!h]
\centering
\includegraphics[height=2.1cm]{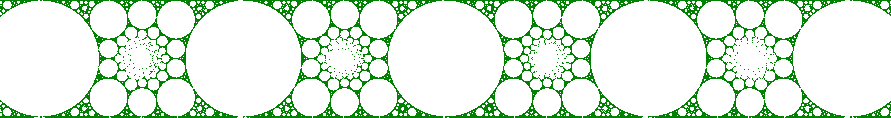}
\end{figure}

\noindent And now the pseudo-code for cancellation tests with regard to the multiplication table.

\begin{extmdframed}
\begin{Verbatim}[fontsize=\footnotesize,frame=none,numbers=left]
//<sub-routine-#1.2:multiplication-table-test>
function __multiplication-table-test__( _digitized_word = "" ) {
//the goal is to check for cancellations. The return value is 1 or 0 if found or not
<let a boolean flag and set it to 0>

//we assume that the index has been converted into the new required base
<split-the-digitized-word-into-an-array-of-single-digits->
<get-the-first-digit-in-the-word>
<get a reference pointer to the related row inside the table>

//we prevent to raise conditional if-statement in the loop
<remove the first digit from the word>

//trasverse and explore the Cayley table
<for each digit in the rest of this array> //sequential read
  <get the next-index in the current row at the index
       referred by the current digit>
    <if the next-index refers to a cancellation, then
       (1) set the above flag to 1
       (2) break this loop
       (3) return the flag to skip the processing of the word under consideration
    >
  <otherwise get a reference pointer to the row inside the-table and
    related to the next-index>
<end-of-for-loop>

<return the flag value>
}
\end{Verbatim}
\end{extmdframed}

\begin{figure}[!h]
\centering
\begin{tabular}{cp{1cm}c}
\includegraphics[height=4cm]{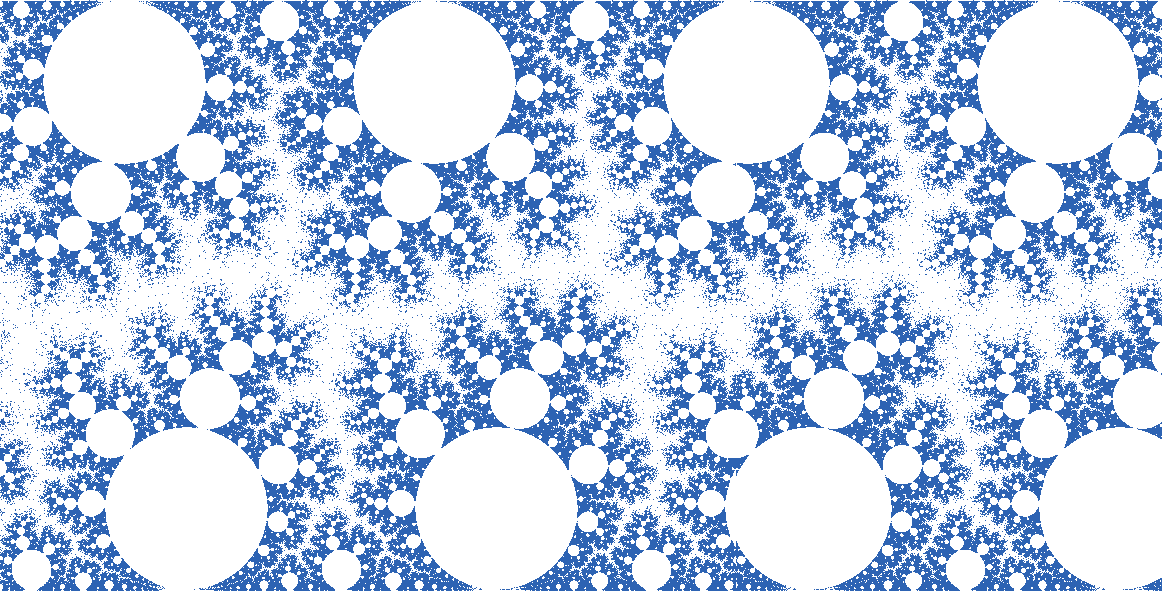}
& &
\includegraphics[height=4cm]{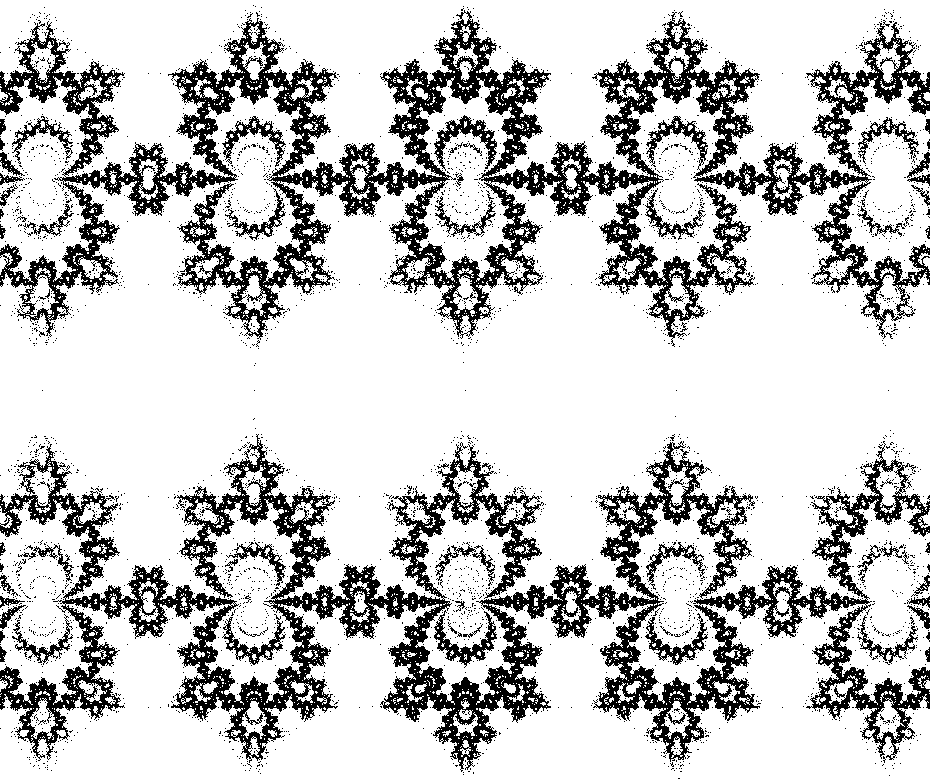}
\end{tabular}
\end{figure} 
\section{Conclusions}\label{section_conclusions}
The \emph{digit}al nature of the index generation algorithm could take away some of the charm tied to the theory of word processing performed by the lexicographic approach (refer to the algebraic theory of commutators at \cite[p. 168]{Indras-2002}) and emanating from the related literature (\cite{Epstein-1992} über alles). Anyway, for practical purposes, this new approach gives the benefit of generating every chain of generators in one only step, much faster than the lexicographic approach. \emph{The transformational character of the index generation algorithm relies upon the simpler and quicker generation of words, no longer coming from the constructive progression, like in the lexicographic approach} which sets up a tortuous track disseminated by the technical drawbacks here discussed at \S \ref{section_drawbacks}, such as appending symbols, checking words and storing them. The necessary computational costs of the index generation algorithm concern cancellation rules and tests.

One more drawback of lexicographic approach concerns the sequential building of words of $n$ symbols, which are deduced from those of length $n-1$. At this regard, we observe that the intrinsic tree structure involves nodes dependency, which demands to start from the root and \emph{walk through} the consecutive nodes in order to get to the given depth. On the contrary, the index generation algorithm enjoys the benefits of numerical sequences which is based upon, allowing to \emph{start}, \emph{stop} and \emph{resume} the generation of strings at any arbitrary element of the sequence. Now we could \emph{jump} from end to end here, instead of \emph{walking through} the interval. It is known that $d=\displaystyle\Bigl\lfloor{\frac{\log(i)}{\log(n)}}\Bigr\rfloor$ returns the number $d$ of digits required to convert $i$ from base 10 to base $n$; so we can explore limit sets inside some interval of integer values, which match with words of length $l$, given $d\leq l\leq D$ for instance.

The author has developed a web application that implements both the lexicographic and the index generation algorithm at {\footnotesize\url{http://alessandrorosa.altervista.org/circles/}}; a number of demos can be run for introductory purposes, or groups be built either geometrically through inversion circles or algebraically via input of arbitrary coefficients into Möbius maps. Refer to \cite{Rosa-2023} for related examples.

\begin{figure}[!h]
\centering
\includegraphics[height=3.3cm]{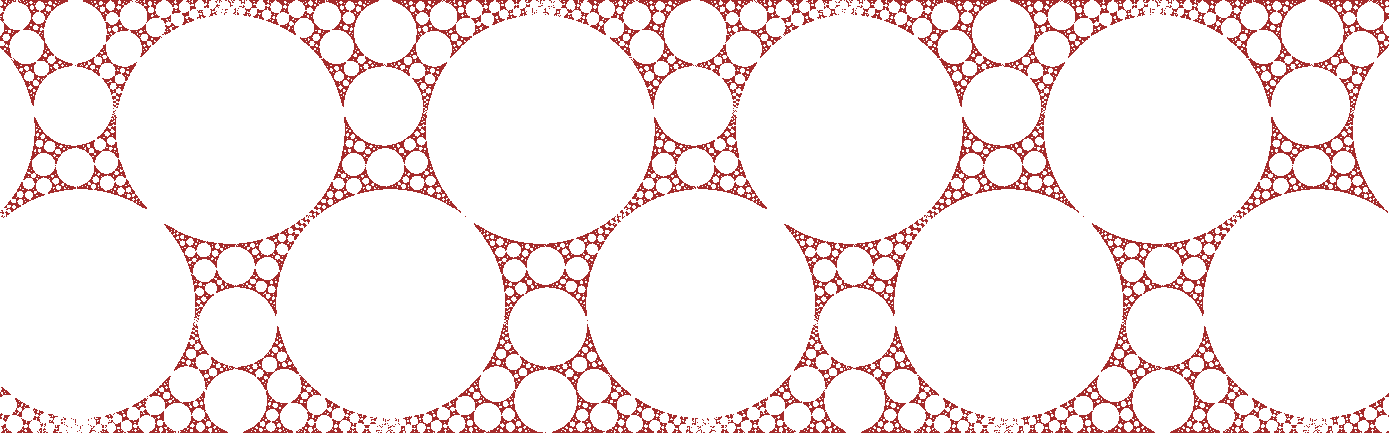}
\end{figure}


\bibliographystyle{amsplain}

\end{document}